# Galaxy formation and chemical evolution

S. Sahijpal

Department of Physics, Panjab University, Chandigarh, India
Email: sandeep@pu.ac.in

**ABSTRACT**

The manner the galaxy accretes matter along with the star formation rates at different epochs, influence the evolution of the stable isotopic inventories of the galaxy. A detailed analysis is presented here to study the dependence of the galactic chemical evolution on the accretion scenario of the galaxy along with the star formation rate during the early accretionary phase of the galactic thick disk and thin disk. Our results indicate that a rapid early accretion of the galaxy during the formation of the galactic thick disk along with an enhanced star formation rate in the early stages of the galaxy accretion could explain the majority of the galactic chemical evolution trends of the major elements. Further, we corroborate the recent suggestions regarding the formation of a massive galactic thick disk rather than the earlier assumed low mass thick disk.

## 1. Introduction

The big-bang origin of the universe occurred ~13.7 billion years (Gyr) ago. The onset of the accretion of the Milky-way galaxy along with the formation of the earliest generations of stars is considered to have commenced quite early. The various stellar nucleosynthetic processes that operate within the numerous generations of stars formed over the evolution of the galaxy established the galactic inventories of all the stable nuclides [1-5]. The gradual growth of the stable isotopic abundances of the galaxy critically depend upon the accretional growth scenario of the galaxy and the star formation rate at different epochs [1-5]. The traditional adopted approach for the formation of the Milky-way galaxy involves episodes of two exponentially declining accretionary growth related with the formation of the galactic thick and thin disks [1, 4, 5]. According to this scenario, the accretional growth of the galactic thick disk occurred over a timescale of the initial ~1 billion years, whereas, the accretional growth of the galactic thin disk occurred over a prolonged timescale of ~7 billion years. This scenario deals with the formation of a massive thin disk with the final surface mass density of ~44 $M_\odot$ $pc^{-2}$ subsequent to the accretion of the galactic thick disk with a final surface mass density of ~10 $M_\odot$ $pc^{-2}$ [see e.g. 4]. This adopted accretion scenario of the galaxy has been used in the previous GCE (galactic chemical evolution) models [1,3-5] to explain the majority of the observed features of the galaxy in terms of the elemental abundance evolution and the supernovae rates. An alternative scenario involving two main accretion episodes with both Gaussian infall rates has been also proposed for the formation of the galaxy [2]. The timescale for the accretion of the galactic thin disk was assumed to be ~4 Gyr in some of these models.

On the basis of the observed elemental composition of the F, G and K dwarf stars in the solar vicinity, two alternative scenarios for the formation of the galaxy have been recently proposed [6-8]. One of these scenarios involve episodes of three exponentially declining accretionary growth [6] involving the rapid accretion of the galactic halo and the thick disk over a timescale of ~0.2 and ~1.25 Gyr, respectively, followed by a prolonged accretion of the thin disk over a timescale of ~6 Gyr. On contrary to the traditional scenarios with a thick disk of low mass (~10 $M_\odot$ $pc^{-2}$) in comparison to a massive thin disk (~44 $M_\odot$ $pc^{-2}$) [1-5], the surface mass density of the thick and thin disks were optimized to ~24 and 30 $M_\odot$ $pc^{-2}$, respectively, in order to explain the elemental abun-dance evolutionary trends [6]. The star formation rate (SFR) in this scenario was evaluated on the basis of the prevailing total surface mass and gas densities [1-5].

An alternative scenario for the formation and the evolution of the galaxy has been recently proposed [7, 8] on the basis of the distribution of [Si/Fe] versus age of the F, G and K dwarf stars in the solar vicinity. This scenario is based on the closed box approximation for the GCE with no inflow or outflow of matter. The galaxy is assumed to have accreted matter rapidly in the early universe. A phase transition between the galactic thick and thin disk occurs ~8-9 Gyr ago. The thick disk acquires almost half of the mass. On contrary to all the previous works on GCE [1-6], the star formation rate was not estimated on the basis of the prevailing total and gas surface mass densities [7, 8]. It was rather considered as a fitting parameter to reproduce the distribution of [Si/Fe] versus age of the F, G and K dwarf stars in the solar vicinity. This work suggests the presence of large amounts of gas in the beginning of the formation of the galaxy and an enhanced early star formation rate.

Based on these two recent proposed scenarios [6-8], in the present work, we study the influence of the rapid early accretion and rapid early star formation history of the galaxy within the framework of episodes of two exponentially declining accretionary growth of the galactic thick and thin disks. We present results on the basis of our N-body numerical simulations of the evolution of the galaxy, whereby, we treat the formation and evolution of several generations of stars during the galactic evolution. The analysis was performed to reproduce the bulk elemental abundance of the sun [5, 9] at the time of its formation. The numerically deduced normalized elemental abundance distribution trends of some major elements were compared with the observed elemental abundance distribution of F, G and K dwarf stars in the solar vicinity. The essential aim is to parametrically impose constraint on the formation of the galactic thick and thin disks along with the star formation history.

## 2. Numerical simulations

The conventional approach of developing the GCE model deals with numerically solving the integro-differential equations dealing with the isotopic abundance evolution [1-4, 6]. An alternative approach of the N-body numerical simulation was developed [5, 10-13] to evolve the galaxy in terms of formation and evolution of numerically simulated stars of different generations. A detailed parametric analysis of the GCE model was performed to constrain the galaxy formation scenario, the star formation rate, the stellar initial mass function and the supernova Ia (SN Ia) contributions to the GCE [5]. The analysis on the influence of the star formation rate and the SN Ia rate on GCE [10] indicated an enhanced star formation rate in the beginning of the formation of the galaxy. Further, it was also concluded [10] that a proposed normalized SN Ia delay distribution [14] corresponding to single degenerate SN Ia with a ~13 % contributions of the prompt SN Ia would adequately explain the SN Ia contributions to GCE. The inhomogeneous GCE models [11] further supported these deductions. A detailed analysis of SN II + SN Ib/c contributions on GCE indicated several limitations related with the use of specific set of stellar nucleosynthetic yields on the GCE models [12].

In the present work, we specifically study the role of the galaxy formation scenario and the star formation rate on GCE. The work is motivated by the recent developments [6-8] that present two distinct views regarding the formation of the galaxy. One of these works [6] deals with the prolonged formation of the galaxy [6] and the other one deal with the rapid accretion of the galaxy in the early universe [7, 8]. However, as mentioned earlier, both of these works indicate rapid formation of a massive thick disk.

**Table 1. The assumed parameters related with the formation of the galactic thick and thin disks.**

| Model | Final accreted surface mass density ($M_\odot$ pc$^{-2}$) | | Simulation fitting parameters |
|---|---|---|---|
| | *Thick disk* | *Thin disk* | (x, f)* |
| Model A | 10 | 44 | (1.52, 0.04) |
| Model B | 30 | 24 | (1.57, 0.03) |
| Model C | 44 | 10 | (1.63, 0.02) |
| Model D | 30 | 24 | (1.56, 0.03) |
| Model E | 44 | 10 | (1.62, 0.02) |

The parameter 'υ' [10] related with the star formation rate during the initial 1 Gyr was assumed to be 1 and 3, respectively, in the case of Models A-C and Models D-E.
* 'x' is the power exponent of the stellar initial mass function in the mass rages 11-100 $M_\odot$ [10-13]
'f' is the fraction of stars in the mass rages 3-16 $M_\odot$ that evolve as binary systems to produce SN Ia

We have developed numerical simulations of GCE in the solar neighborhood using the formulation discussed elaborately in the earlier works [5, 10-13]. We have specifically adopted our recent models for the formulation of the SN Ia and the star formation rate [10]. We here avoid repetitive discussion of the details of the numerical simulations. We adopted the scenario related with two exponentially declining accretionary growth episodes associated with the formation of the galactic thick and thin disks with the assumed timescales of 1 Gyr and 7 Gyr, respectively [1, 3-5]. The extent of the distribution of mass among the galactic thick and thin disks is treated as one of the simulation parameters to understand its dependence on GCE (**Table 1**). The traditional galaxy accretion scenario deals with the formation of a galactic thick disk with a final surface mass density of ~10 $M_\odot$ pc$^{-2}$ (**Table 1**). This is subsequently followed by the accretion of a massive thin disk with a final surface mass density of ~44 $M_\odot$ pc$^{-2}$ [see e.g. 4]. The Model A (**Table 1**) refers to this specific scenario. We ran Model B & C by varying the final accreted mass distribution of the galactic thick and thin disks. The star formation rate in the present work was estimated on the basis of the prevailing total surface and gas baryonic matter density of the galaxy [4, 5, 10, 11]. This approach is identical to the recent work [6]. The Models A, B and C were run by assuming a value of 1 for the star formation rate parameter, 'υ' (see Equation 1 of [10]). The simulations, Model D and E were run with a value of 3 for υ. It has been demonstrated earlier [10, 11] that the GCE models with this particular value explain majority of the observed elemental abundances distribution of the F, G and K dwarf stars in the solar vicinity. It should be noted that the Models B and D (**Table 1**) in the present work deals with the accretion of the galactic thick disk of substantial mass that is comparable to the thin disk. These scenarios are almost identical to the recent GCE scenario [6]. However, the Models C and E (**Table 1**) deal with rapid accretion of substantial matter in the early evolution of the galaxy. Although, this scenario does not exactly correspond to the SFR and galaxy formation formulation that was adopted in the recent works [7, 8], the two distinct approaches however would eventually result in an early rapid growth of the galaxy and an early enhanced star formation.

During the evolution of the galaxy in the solar vicinity, the stars in the mass range 0.1-100 $M_\odot$ were formed at the time of formation of stellar clusters of different generations according to an assumed piece-wise three step normalized stellar initial mass function (IMF) [5, 10-13]. The power exponent, x, of the IMF in the mass rages 11-100 $M_\odot$ was considered as one of the simulation parameters (**Table 1**) to reproduce the metallicity of 0.014 [9] and [Fe/H] ~ 0 at the time of formation of the solar system [10-13]. The stellar nucleosynthetic yields of SN II + SNI b/c [15], the low and the intermediate mass stars [16] and SN Ia [17] were used to deduce the stable isotopic abundance evolution of the various elements from hydrogen to zinc over the galactic timescales. We adopted the SN Ia formulation that involves a SN Ia normalized delay distribution [14] corresponding to single degenerate SN Ia with a ~13 % contributions of the prompt SN Ia [10]. A fraction, f, of stars in the mass range 3-16 $M_\odot$ from the IMF was considered as one of the other simulation parameters apart from the parameter, x (**Table 1**). This fraction of stars evolves as binary stars and eventually produces SN Ia, the main contributors of the iron-peaked nuclides to the galaxy. The optimized values of the simulation fitting parameters are mentioned in **Table 1**.

## 3. Results and Discussion

Numerical simulations were run to study the dependence of the galaxy formation scenario and the star formation rate on the galactic elemental abundance evolution (**Table 1**). The star formation rates (SFR) inferred from the simulations are presented in **Figure 1**. Compared to Model A, the Models B and C infer higher star formation rate during the early evolution of the galaxy. The star formation rate parameter, $\upsilon$ (see Equation 1 of [10]) in these models was assumed to be 1. It is interesting that the deduced SFR in the case of Model C follows almost an identical trend as observed in the recent work based on treating the SFR as a fitting parameter [7]. As mentioned earlier, even though, the present adopted approach is distinct from the recent work [7], the two approaches however identically invoke an early rapid accretion of the galaxy and a high SFR. An increase in the value of SFR parameter, $\upsilon$ to 3 results in a further enhancement of the SFR as is noticed in the case of simulations, Model D and Model E. This aspect was also noticed in the recent work [6]. It should be mentioned that the galactic thick disk is assumed to accrete over a timescale of 1 Gyr, and the parameter $\upsilon$ is taken to be unity during the accretion of galactic thin disk in all the simulations. The enhanced SRF during the early accretion of the galaxy results in a lower SRF during the latter evolutionary stages of the galaxy in all the simulations.

The numerically simulated total baryonic surface matter density accretion and distribution for the various models is presented in **Figure 2**. Due to substantially more baryonic matter accretion during the galactic thick disk accretionary stage (**Table 1**), the Models B and D acquire an earlier high matter accretion compared to the Model A. The Models C and E result in an even higher matter accretion during the early evolution of the galaxy. All the models eventually acquire an identical present day observed value of ~54 $M_\odot$ pc$^{-2}$ for the combined galactic thick and thin disks total baryonic matter surface density. The distributions of the baryonic matter among stars (+ stellar remnants) and interstellar gas are also presented in **Figure 2** for the various models. Compared to the Model A, the early enhanced star formation in the models B and C occurs due to the early rapid accretion of the galaxy. There is a sharp rise in the surface gas density prior to enhanced star formation. Compared to the Models B and C, in the case of the Models D and E, a further enhancement in the SFR during the initial couple of billion years is due to the choice of a higher value of 3 for the SFR parameter, $\upsilon$ (**Table 1**). However, it should be noted that this does not significantly alter the SFR during the latter stages of the galaxy evolution.

The predicted supernovae rates for the various simulations are presented in **Figure 3**. The SNII +SNIb/c supernovae rates follow the star formation rate temporal trends (**Figure 1**). This is due to the fact that most of the massive (>11 $M_\odot$) stars explode as SNII +SNIb/c within a timescale of ~25 million years (Myr.) subsequent to their formation in a stellar cluster [11-13]. The enhancement in the star formation rate (**Figure 1**) in the earlier stages of galaxy results in earlier enhanced SNII +SNIb/c rates. This would result in an earlier enrichment of the stellar nucleosynthetic yields of the massive stars. It should be also noted that the power exponent, x, of the stellar initial mass function increases due to the earlier rapid accretion of the galaxy or the enhancement in the SFR (**Table 1**). Further, as mentioned earlier the SN Ia rates are according to the normalized delay distribution of single degenerate SN Ia with a ~13 % contributions of the prompt SN Ia [10, 14]. The fraction, f, of the stars in the mass range 3-16 $M_\odot$ that eventually results in SN Ia decreases with the earlier rapid accretion of the galaxy or the enhancement of SFR (**Table 1**).

The estimated evolutionary trends in the [Fe/H] versus age are presented in **Figure 4** for all the models. As far as the earlier temporal evolutionary trend in the [Fe/H] are concerned, the Model D explains the best match with the observational data of the F, G and K dwarf stars in the solar vicinity. The Model D corresponds to the scenario dealing with the early rapid accretion of a galactic thick disk with a mass comparable to the galactic thin disk (**Table 1**). This scenario also deals with an enhanced star formation rate with a value of 3 for the SFR parameter, $\upsilon$ [10-11]. Based on the comparison among the Models B-C and D-E (**Figure 4**), it should be noted that a further increase in the mass of the galactic thick disk in comparison with the thin disk (**Table 1**) makes the mismatch larger with the observational data in the earlier temporal evolutionary trends in [Fe/H].

The temporal normalized evolutionary trends of some major elements for the Models B, C, D and E are presented in **Figure 5**. As mentioned earlier an enhanced SNII +SNIb/c rates in the early evolution of the galaxy result in an earlier enrichment of heavier elements due to the stellar nucleosynthetic yields of massive stars. Except for nitrogen, the Models D and E with an enhanced SFR in the early galaxy results in earlier high elemental yields compared to the Models B and C. The Models D and E provide a better match with the observational data. Further, except for carbon, nitrogen, copper and zinc, the elemental evolutionary trends among the Models B-C and D-E, having an identical value of the SFR parameter, $\upsilon$, are not distinct. Based on the normalized [Si/Fe] versus [Fe/H] (**Figure 5**) and [Fe/H] versus age (**Figure 4**), it seems that the Model D could be most preferred choice for [Fe/H] > -1.0 [7]. Finally, as discussed in the earlier works it is impossible to match all the elemental evolutionary trends with the observational data [5, 10-12].

## 4. Conclusions

Based on the detailed numerical simulations of the galactic chemical evolution, we corroborate the recent findings [6, 7] that the formation of the galactic thick disk of mass comparable to the mass of the galactic thin disk occurred rapidly in the early universe. This was followed by the accretion of a galactic thin disk perhaps over a prolonged timescale. Further, we present predicted normalized elemental evolutionary trends for several elements for the first time.

## 5. Acknowledgements

This work is supported by a PLANEX (ISRO) grant

### REFERENCES

[1] Chiappini, C., Matteucci, F. and Gratton, R., (1997) The chemical evolution of the galaxy: The two-Infall model. *The Astrophysical Journal*, **477**, 765-780.

[2] Chang, R. X., Hou, J. L., Shu C. G. and Fu, C. Q. (1999) Two-component model for the chemical evolution of the Galactic disk. *Astronomy and Astrophysics*, **350**, 38-48.

[3] Tutukov, A. V., Shustov, B. M. and Wiebe, D. S. (2000) The stellar epoch in the evolution of the galaxy. *Astronomy Reports*, **44**, 711-718.

[4] Alibés, A., Labay, J. and Canal, R. (2001) Galactic chemical abundance evolution in the solar neighborhood up to the iron peak. *Astronomy and Astrophysics*, **370**, 1103-1121.

[5] Sahijpal, S. and Gupta, G. (2013) Numerical simulation of the galactic chemical evolution: The revised solar abundance. *Meteoritics and Planetary Science Journal*, **48**, 1007-1033.

[6] Micali, A., Matteucci, F. and Romano, D. (2013) The chemical evolution of the Milky Way: The three infall model. *Monthly Notices of the Royal Astronomical*, **436**, 1648-1658.

[7] Snaith, O. N. et al. (2014) The dominant epoch of star formation in the milky way formed the thick disk. *The Astrophysical Journal Letters*, **781**, article id. L31.

[8] Haywood, M. (2014) Galactic chemical evolution revisited. *Memorie della Societa Astronomica Italiana Supplement*, eprint arXiv:1401.1864

[9] Asplund, M., Grevesse, N., Sauval, A. J. and Scott, P. (2009) The chemical composition of the Sun. *Annual Review of Astronomy and Astrophysics*, **47**, 481-522.

[10] Sahijpal, S. (2013) Influence of supernova SN Ia rate and the early star formation rate on the galactic chemical evolution. *International Journal of Astrophysics and Astronomy*, **3**, 344-352.

[11] Sahijpal, S. (2013) Inhomogeneous chemical evolution of the galaxy in the solar neighbourhood. *Journal of Astrophysics and Astronomy*, **34**, 297-316.

[12] Sahijpal, S. (2014) Contributions of type II and Ib/c supernovae to Galactic chemical evolution. *Research in Astronomy and Astrophysics*, **14**, 693-704.

[13] Sahijpal, S. (2014) Evolution of the galaxy and the birth of the solar system: The short-lived nuclides connection. *Journal of Astrophysics and Astronomy*, **35**, 121-142.

[14] Matteucci, F. Spitoni, E., Recchi, S. and Valiante R. (2009) The effect of different type Ia supernova progenitors on Galactic chemical evolution. *Astronomy and Astrophysics*, **501**, 531-538.

[15] Woosley, S. E. and Weaver, T. A. (1995) The Evolution and Explosion of Massive Stars. II. Explosive Hydrodynamics and Nucleosynthesis. *The Astrophysical Journal Supplement*, **101**, 181-235.

[16] Karakas, A. I. and Lattanzio, (2007) Stellar models and yields of asymptotic giant branch stars. *Publications of the Astronomical Society of Australia*, **24**, 103-117.

[17] Iwamoto, K. et al. (1999) Nucleosynthesis in Chandrasekhar mass models for type Ia supernovae and constraints on progenitor systems and burning-front propagation. *The Astrophysical Journal Supplement*, **125**, 439-462.

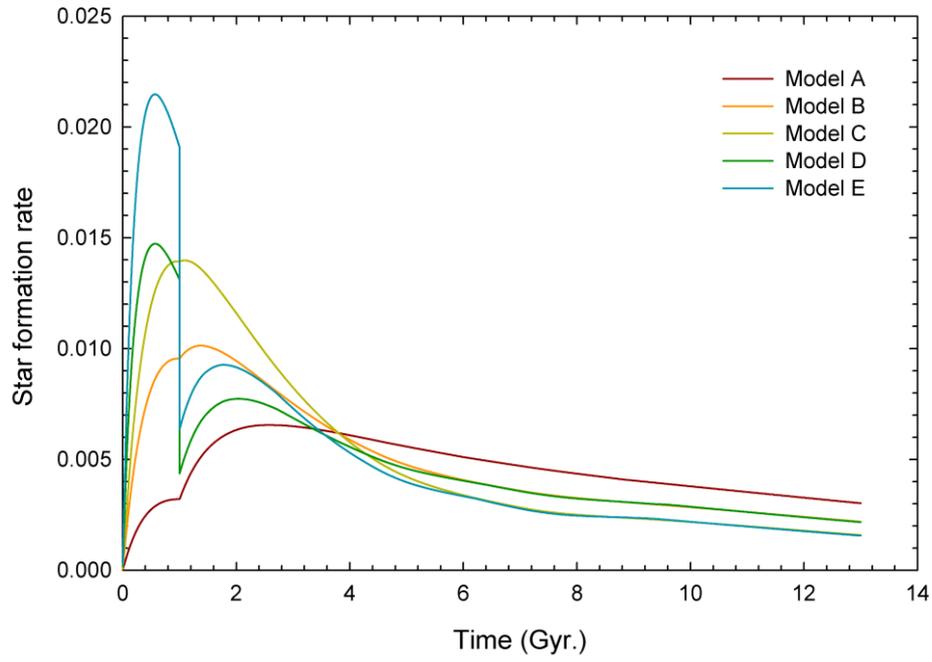

**Figure 1.**

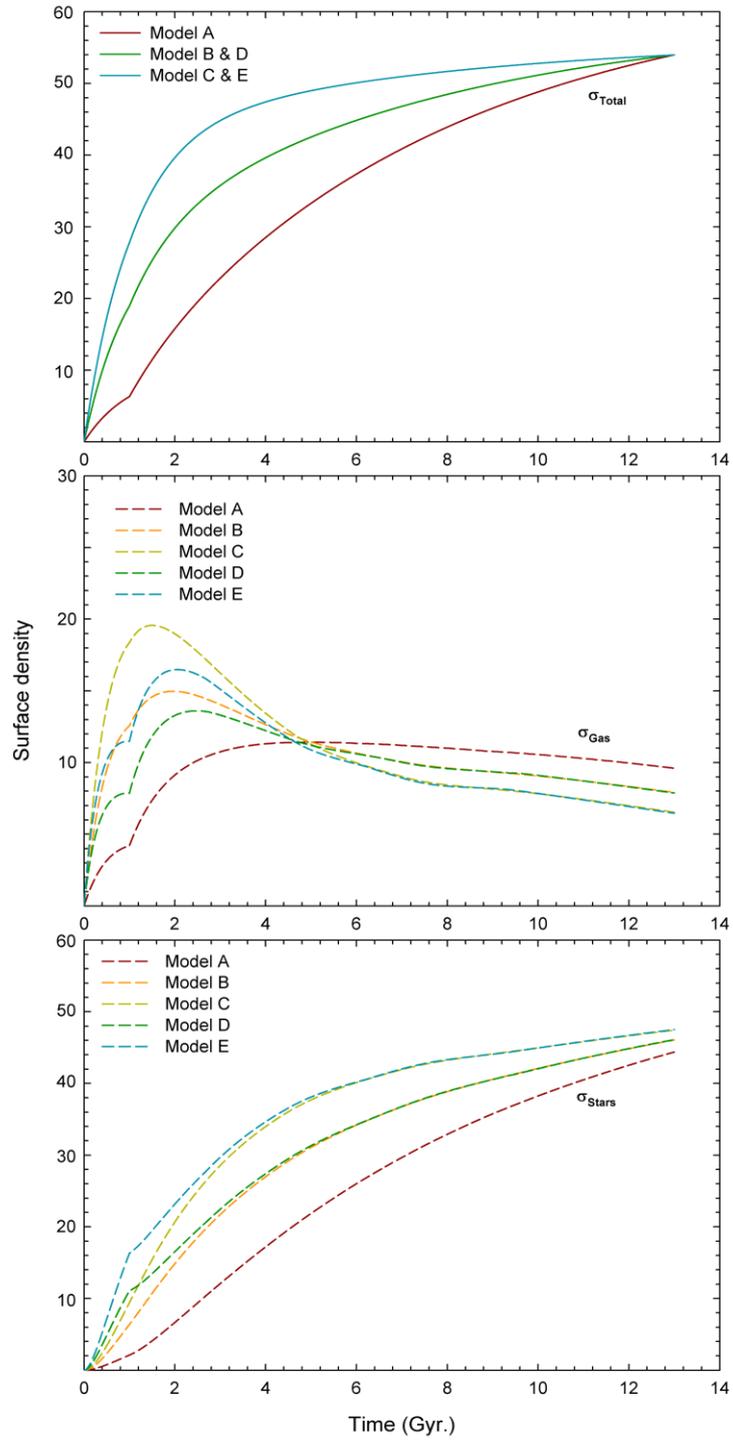

**Figure 2.**

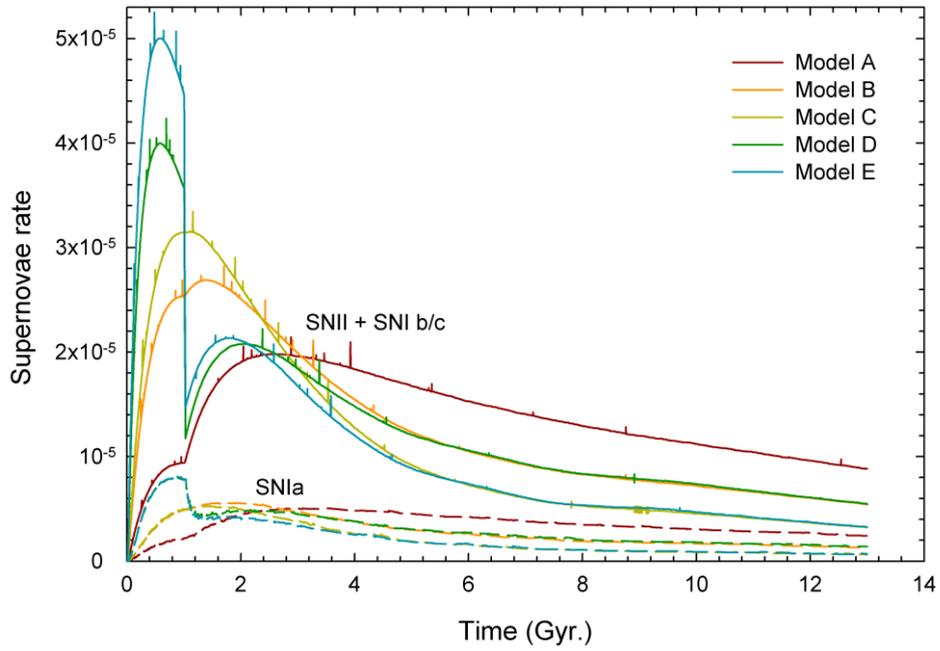

**Figure 3.**

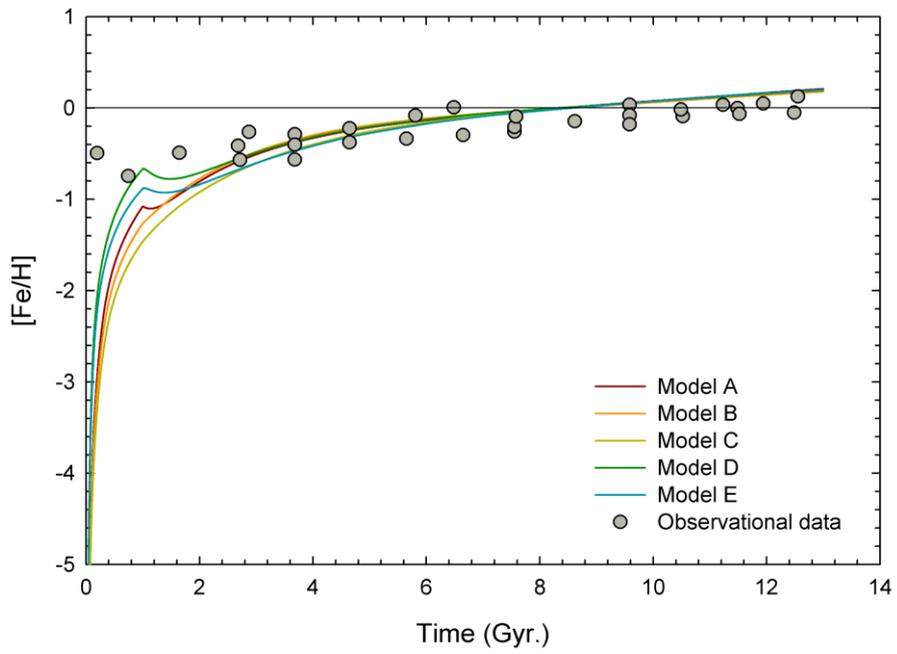

**Figure 4.**

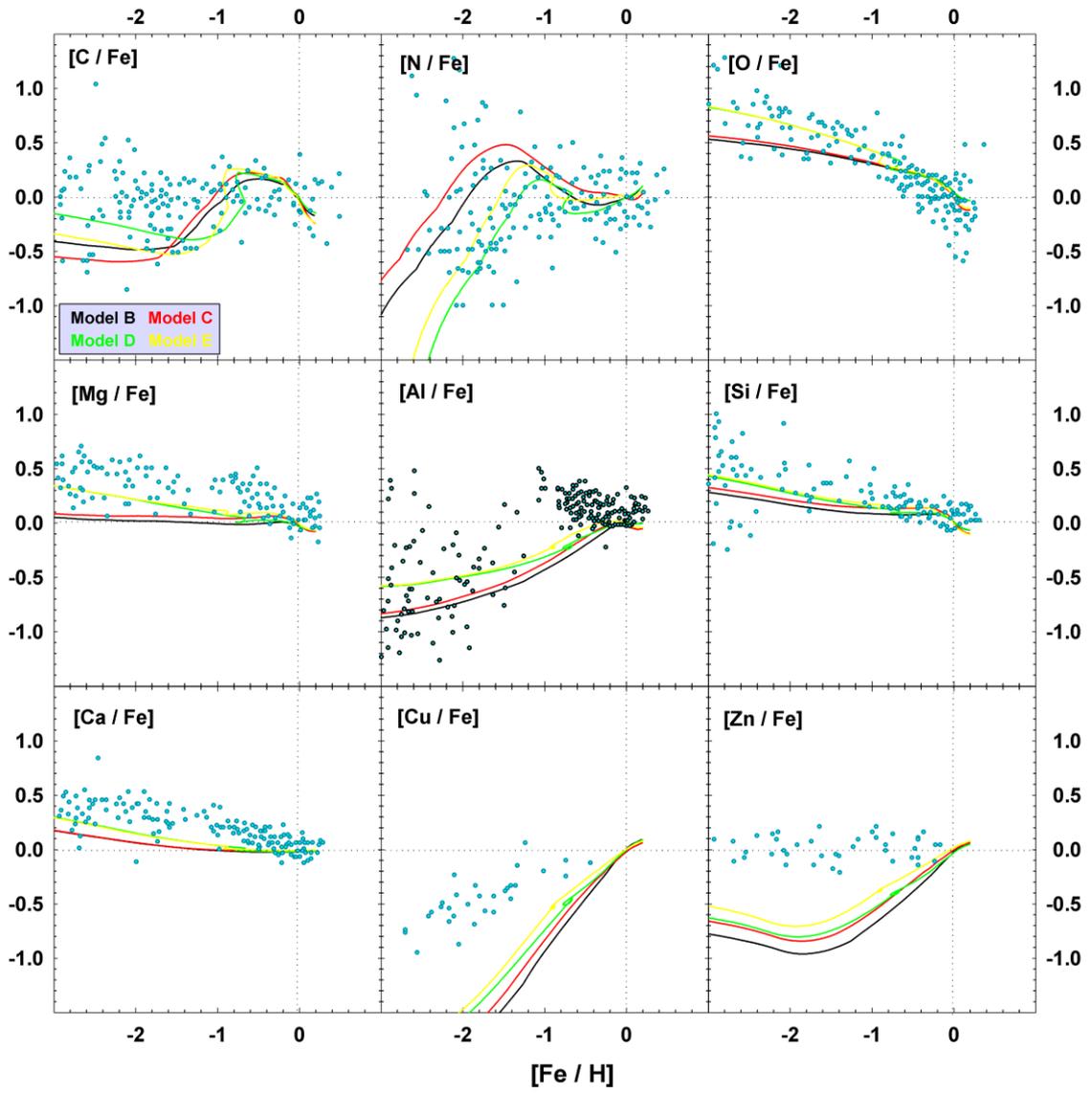

**Figure 5.**

**Figure Caption**

Figure 1. The star formation rate (SFR, $M_\odot$ pc$^{-2}$ Myr$^{-1}$) for the various GCE models.

Figure 2. The estimated total baryonic surface mass density ($M_\odot$ pc$^{-2}$), the gas mass density, the stellar (+ stellar remnant) mass density, for the GCE models.

Figure 3. The estimated supernova rates (in pc pc$^{-2}$ Myr$^{-1}$) for the various GCE models.

Figure 4. The estimated evolution of [Fe/H] for the GCE models. The observational data of the F, G and K dwarf stars in the solar vicinity is also presented [4].

Figure 5. The estimated normalized elemental abundance evolution for the various simulations. The observational data of the F, G and K dwarf stars in the solar vicinity is also presented [5].